\documentclass[conference]{IEEEtran}
\IEEEoverridecommandlockouts
\usepackage{cite}
\usepackage{amsmath,amssymb,amsfonts}
\usepackage{algorithm} 
\usepackage{algpseudocode} 
\usepackage{graphicx}
\usepackage{textcomp}
\usepackage{xcolor}
\usepackage{adjustbox}
\usepackage{booktabs}
\usepackage{float}

\let\bf\bfseries

\def\BibTeX{{\rm B\kern-.05em{\sc i\kern-.025em b}\kern-.08em
    T\kern-.1667em\lower.7ex\hbox{E}\kern-.125emX}}
\begin{document}

\title{AMA-K: Aggressive Multi-Temporal Allocation An Algorithm for Aggressive Online Portfolio Selection\\
}

\author{\IEEEauthorblockN{Matthew Kruger }
\IEEEauthorblockA{\textit{Computer Science and Applied Mathematics} \\
\textit{University of the Witwatersrand}\\
Johannesburg, South Africa \\
}
\and
\IEEEauthorblockN{Terence L. van Zyl}
\IEEEauthorblockA{\textit{Institute for Intelligent Systems} \\
\textit{University of Johannesburg}\\
Johannesburg, South Africa \\
}
\and
\IEEEauthorblockN{Andrew Paskaramoorthy}
\IEEEauthorblockA{\textit{Computer Science and Applied Mathematics} \\
\textit{University of the Witwatersrand}\\
Johannesburg, South Africa \\
}

}

\maketitle

\begin{abstract}
Online portfolio selection is an integral component of wealth management. The fundamental undertaking is to maximise returns while minimising risk given investor constraints.
We aim to examine and improve modern strategies to generate higher returns in a variety of market conditions.
By integrating simple data mining, optimisation techniques and machine learning procedures, we aim to generate aggressive and consistent high yield portfolios. This leads to a new methodology of Pattern-Matching that may yield further advances in dynamic and competitive portfolio construction.
The resulting strategies outperform a variety of benchmarks, when compared using Maximum Drawdown, Annualised Percentage Yield and Annualised Sharpe Ratio, that make use of similar approaches. The proposed strategy returns showcase acceptable risk with high reward that performs well in a variety of market conditions.
We conclude that our algorithm provides an improvement in searching for optimal portfolios compared to existing methods.
\end{abstract}


\section{Introduction}

Multi-period portfolio choice is a central problem in finance. It is described by an investor who faces the problem of determining how to sequentially allocate his capital to maximise some performance measure over multiple periods. Online portfolio selection algorithms tackle the problem of maximising cumulative wealth by adaptively identifying and exploiting patterns in historical data~\cite{Li}. The key feature of these algorithms is that they are \textit{online}: patterns and portfolio decisions update upon the arrival of new data, thereby adapting to changing market conditions~\cite{Li}.


Online portfolio selection algorithms can be classified according to their update scheme. Traditional algorithms forecast asset returns and are used to update the current portfolio. Li \textit{et al.} (2014) classifies traditional algorithms into the following categories \cite{Li}: 
\begin{itemize}
    \item Follow-The-Winner (FTR) algorithms assume that recent stock performance would persist and so transfer capital from the worst-performing stocks to the best-performing stocks.
    \item Follow-The-Loser (FTL) algorithms assume that recent stock performance would revert to a long-run mean and so transfer capital from the best-performing stocks to the worst-performing stocks.
    \item Pattern-Matching based (PM) algorithms assume that market conditions repeat themselves and so they allocate capital based on what was optimal for similar historical periods.
\end{itemize}

The Pattern-Matching based approach has the least restrictive assumption about market behaviour. This affords greater flexibility in algorithm design and allows these algorithms to exploit a wider range of market conditions, thereby outperforming the other approaches \cite{Gyorfi, CORN, RACORN, DRICORN}. In particular, the CORN-K (CORrelation-driven Non-parametric learning) algorithm appears to demonstrate the best results. Recently, the CORN-K algorithm has been extended to incorporate risk in its portfolio selection \cite{RACORN, DRICORN}.

However, the CORN-K algorithm (and its extensions) often output a cautious portfolio which restrict its returns. In short, this occurs when the algorithm is unable to detect a subset of historical data that is similar to the recent data and therefore, allocates wealth equally across assets. 

To do this, we propose the \textbf{A}ggressive \textbf{M}ulti-Temporal \textbf{A}llocation (AMA-K) algorithm, which combines the Pattern-Matching and Follow-the-Winner principles.

The rest of this paper is organised as follows. In Section \ref{opsSection}, we define the portfolio selection problem. Section \ref{cornSection} discusses the CORN-K algorithm.  Section \ref{amakSection} introduces our AMA-K algorithm. 

\section{Online Portfolio Selection} \label{opsSection}
An investor wants to allocate his initial capital $S_0$ into a portfolio of $m$ securities for each of the $n$ trading days to maximise his terminal wealth $S_n$. The investor's portfolios are represented
by $\textbf{b}_{t} = (b_{(t,1)}, \dots , b_{(t,m)})$, where $b_{(t,j)}$ is a proportion of the capital invested in security $j$ at time $t$. Furthermore, portfolio positions are constrained to be non-negative $b_{(t,j)} \geq 0$ and all capital is invested at each period $\sum_{j=1}^{m}b_{(t,j)} = 1$.

Define the price relative for security $j \in 1, \dots, m$ at day $t \in 1, \dots, n$ as $x_{(t,j)} = \frac{P(t,j)}{P(t-1,j)}$, where $P(t,j)$ denotes the log-price. Hence, denote the price relative vector as $\mathbf{x}_{t} = (x_{(t,1)}, \dots , x_{(t,m)}) \in \mathbb{R}^{m}_{+}$. A sequence of price relative vectors are used to define a market window $\textbf{X}^{t}_{t-w} = (\textbf{x}_{t-w}, \textbf{x}_{t-w+1}, ... ,\textbf{x}_{t})$, where $w$ is the given window size.

An online portfolio selection algorithm $\mathcal{A}$ is a function that takes the historical price data at time $t$ and outputs a portfolio:
\begin{equation}
    \mathcal{A}\left(\cup_{i=1}^{t-1} \lbrace \textbf{X}^{t-1}_{i} \rbrace \right) = \textbf{b}_{t}.
\end{equation}
The portfolio is constructed at the start of period $t$, using all information up until then. The terminal wealth at the end of period $n$ is given by:
\begin{equation}
    S_n = S_0\sum_{t=0}^{n-1} (\mathbf{b}_{t+1}'\mathbf{x}_{t+1}).
\end{equation}

For tractability, we make the following assumptions: each asset is arbitrarily divisible, desired quantities can be traded at the most recent closing price, and market prices are not affected by the investor's actions. In addition, we ignore trading costs and do not allow for borrowing or short-selling.  
\section{Correlation-driven Non-parametric Learning}\label{cornSection}
CORN-Based strategies use experts that construct portfolios using previous market windows.
Experts have a portfolio $\mathbf{b}_{t}$ and  have a cumulative wealth $s_{t}$, at time $t$. $P \times W$ experts are considered, each defined by a window size $w \in (1, \ldots , W$) and a \textit{Pearson product-moment correlation coefficient} threshold $\rho \in (\frac{1}{P},\ldots,\frac{P-1}{P}$)~\cite{CORN}. 
In top $K$ based strategies, the $K$ experts with the most cumulative wealth at time $t$ have their portfolios combined. Each expert is responsible for $\frac{1}{K}$ wealth of the portfolio's allocation for a given day. This combined portfolio is the agent's portfolio at time $t$.

Each expert compares the most recent market window at time $t$ with all historical market windows of the same size. Each expert searches for their optimal portfolio using a given set of data that is equal to or greater than their respective $\rho$. Days that match this required $\rho$ are called \textit{correlation similar days} and is represented by $C_{t}(w,p)$~\cite{CORN}. Experts update their wealth at the end of a day using this portfolio and the day's returns.

An expert's portfolio is determined by $\textbf{b}$ that maximises Equation~\ref{Eq:cornMaximisation} at time $t$:
\begin{equation}
    \mathbf{b}_{t}(w,p) = \operatorname*{argmax}_\mathbf{b} \prod_{i\in C_{t}(w,p)} (\mathbf{b \cdot x}_{i})
    \label{Eq:cornMaximisation}
\end{equation}

At times, the correlation similar set of days may be small or empty. In this case the expert returns uniform portfolio~\cite{CORN,RACORN,DRICORN}. A uniform portfolio is when wealth is equally distributed amongst all assets - which generally have lower returns. DRICORN-K is a variation of CORN-K that classifies the market and adjusts accordingly. Classification is done through the use of $\beta$ (market Beta) in searching for the optimal portfolio ~\cite{DRICORN}. Utilising $\beta$ allows for more aggressive/defensive portfolios based on current market conditions~\cite{DRICORN}. At times, $\beta$ does not impact the portfolio construction, in which case DRICORN-K returns a similar portfolio to CORN-K.


\section{Method}\label{amakSection}


Clustering previously has been employed in online portfolio selection~\cite{kmeansTaiwan,kMeansBombay,iranianClustering,clusteringItaly, nanda}. Similar approaches are given by Khedmati \textit{et al.} (2020) where portfolios are optimised using clustering techniques, market windows, Pattern-Matching and \textit{similar day samples}~\cite{iranianClustering, kmeansTaiwan}.
Nanda \textit{et al.} (2010) found that K-means clustering provided the best result for online portfolio selection based on cluster compactness using the Bombay Stock Exchange~\cite{nanda}. Our work extends on these previous successes by directly integrating clustering into the CORN-K framework. Further, we introduce a more effective low dimensional representation for market windows that improve the clustering results. 

A limitation of CORN-K's use of \textit{correlation similar days} is correlation similar days are rare and usually only common amongst experts with smaller window sizes and lower values for $\rho$. Hence, experts that have a suitable quantity of data to produce inference in the market dynamics are unable to do so. To overcome this limitation, we use online K-Means clustering (K$\mu$-online) with Manhattan distance as an alternative to discover sets of \textit{cluster similar days}.
These cluster similar days do not require a correlation coefficient threshold and are considered similar if they belong to the same centroid. Manhattan distance is selected for computational efficiency - alternate metrics could be considered.

Our variation to CORN-K lies in dealing with empty sets of correlation similar days. If we encounter a day where the agent has $C_{t}(w,p) = \emptyset$ and the market window size is $w > 2$, we make use of our current day's (day $t$) market vector's assigned cluster as created in Algorithm~\ref{alg:supClustering}. We let the correlation similar set be all days assigned to the same cluster and maximise using Equation~\ref{Eq:cornMaximisation}.

\subsection{Aggressive Multi-Temporal Allocation}
We have chosen to maintain the method of choosing $K$ best experts in our algorithm. Furthermore, we use the concept of market windows ($\textbf{X}^{t}_{t-w}$ for a market window of size $w$ at day $t$), these are matrices that represent consecutive market days' price relative vectors across all shares.

\subsubsection{Agent Memory}
At the start, all the agent's experts have $d$ days of market history. As the algorithm proceeds, new days are added to the agent's memory until it has $2d$ days of market history. At $2d$ days, the agent forgets all but the most recent $d$ days of market history. The choice of $d$ ensures that the agent considers only recent price movements and the length of $2d$ keeps the agent's portfolio allocations ``stable``. A small $d$ value results in an myopic agent that exploits volatility. A high $d$ value results in an agent that looks for``blue chips`` that represent long-term growth trends.

Each market window $X_{t-w}^t$ is represented by a market vector $m_v \in \mathbb{R}^{4}$ with components:
$\textbf{m}_v=[$ the sum of each stock's mean in the market window (Equation~\ref{eq:marketVectorMean}), the market window's mean, the sum of each stock's variance (Equation~\ref{eq:marketVectorVar}) and the variance of the market window $]$. 

\begin{equation}
    \label{eq:marketVectorMean}
    \frac{1}{w}
    \sum_{j=1}^{m} 
    \sum_{k=t-w}^{t}
    \textbf{X}_{(k,j)}^{(t,j)},
\end{equation}
where $j$ represents the $j^{th}$ asset.
\begin{equation}
    \label{eq:marketVectorVar}
    \frac{1}{w}
    \sum_{j=1}^{m}
    \sum_{k=t-w}^{t} \left(\textbf{X}_{(k,j)}^{(t,j)} - \mu_{j}^{2} \right),
\end{equation}
where $\mu_j$ is the average price relative for the $j^{th}$ asset in the given market window.

We initialise the algorithm with $W=5$, windows of size $w_i$ where $i \in (1,\ldots, W)$ market vectors and $W$ is the maximum window size. $W$'s value is from the original CORN paper~\cite{Li}. We initialise the number of cluster centroids as $\lfloor \frac{d}{3} \rfloor$ ($d$ represents the number of days). The initial centroid amount was determined using the validation data.

Subsequently, we shift the market window forward by one day and assign $W$ new market vectors for their respective windows $w_i$ to their nearest clusters for that agent's experts.
Noting that market vectors ($w_{i})$ are assigned to experts with the same window size.
We re-do the clustering every $\lfloor \frac{d}{3} \rfloor$ days to keep the allocation of market vectors uniform and relevant. The additional centroids allow for the new unseen market vectors to be represented. Even though this readjustment interval was determined using the validation set, the strategy was found to be insensitive to intervals in the range $\lfloor \frac{d}{4} \rfloor$ to $\lfloor \frac{d}{5} \rfloor$.

Since the K$\mu$-online algorithm does not converge, we terminate the clustering when reassignments affect only $\varepsilon =0.6\%$ of market vectors. The $0.6\%$ is a manually determined parameter for the general case when the assignment of market vectors to clusters proceeded normally. If the $0.6\%$ threshold is not obtained within ten re-initialisation attempts, we use the last readjustment. This maximum number of attempts allow the clustering process to proceed which may yield undesirable cluster assignments.

Every $2d$ days we reset the K$\mu$-online clustering using the most recent $d$ days of data and $\lfloor \frac{d}{3} \rfloor$ centroids. The reason why we reset every $2d$ days is that it allows the algorithm to provide a balance between factoring in new information and acting less erratically from continuously switching between asset allocations.

\begin{algorithm}[!htb]
	\caption{Supplementary Clusters} \label{alg:supClustering}
	\begin{algorithmic}[1]
	    \item Input:  $X_{1}^{n} = (\textbf{x}_{1},\ldots,\textbf{x}_{n})$ containing $n$ days of market history, with $W$ the max window size, $d \in (W,n)$ as the selected memory, $c=\lfloor \frac{d}{3} \rfloor$ centroids and $\varepsilon$ tolerance to cluster reassignment.
		\State $\textbf{M}_c$ = [] \Comment{Array of Centroids}
		\State $\textbf{M}_v$ = [] \Comment{Array of Market Vectors}

        \Function{reInit}{$X_{m}^{p}$}
            \For{$w=1$ to $W$} \Comment{Calculate $\textbf{M}_{c}$'s component $\textbf{m}_{c}$}
        	    \State $\textbf{m}_{v}$ = marketVectors($X_{m}^{p}, w$)
        	    \State $\textbf{m}_{c}$ = K$\mu$-online($\textbf{m}_{v}$, $c$, $\epsilon$ , $w$)

        	    \State $ \textbf{M}_{v} = \textbf{M}_{v}\cup \textbf{m}_{v}$
                \State $\textbf{M}_{c} = \textbf{M}_{c} \cup \textbf{m}_{c}$
        	\EndFor
        \EndFunction
    	\State reInit($\textbf{X}^{d}_{1}$)
    	\For{$t = d + 1$ to n} \Comment{Trading for the duration}
    	    \If{$t$ \% $2d == 0$ } \Comment{Re-doing clustering}
    	        \State $c=\lfloor \frac{d}{3} \rfloor$; $\textbf{M}_{v}$ = [];  $\textbf{M}_{c}$ = []
    	        \State reInit($X_{1}^{d}$)
    	   \Else
    	        \If{$t$ \% $\lfloor \frac{d}{3} \rfloor == 0$} \Comment{Adding a centroid}
    	            \State $c = c+1$ \Comment{New empty centroid}
    	        \EndIf  
    	        \State $\textbf{m}_v$ =  marketVectors($X^{t}_{t-d}, w$)
    	        \State $\textbf{M}_{v} = \textbf{M}_{v}\cup \textbf{m}_{v}$ \Comment{Combine market vectors}
            	\State $\textbf{M}_{c}$ = K$\mu$-online($\textbf{M}_{v}$, $c$ ,$\epsilon$ , $w$)
    	   \EndIf
    	\EndFor
    \end{algorithmic}
\end{algorithm}

\subsubsection{AMA-K} 
\label{subsec:AMA-K}
Given that the various (10-day, 120-day and 190-day) agents perform well for their respective time horizons as shown by Table~\ref{tab:effects-of-memory}, we create an algorithm that combines the agents.
This combination creates an agent with three time horizons that we have defined as; short (10-day), medium (120-day) and long (190-day).
Each time horizon has its own set of experts associated to it.
The clustering algorithm is repeated for each sub-agent using their $d$ value.
Here a sub-agent is an agent over a specific $d$-day time horizon.
We take the portfolio for each time horizon and normalize it such that it represents the proportion of the sub-agent's wealth allocated to each asset at day $t$.
These portfolios are merged. The resulting portfolio is then divided by the number of time horizons under consideration (here we divide by three).
This resultant portfolio may have a diverse range of assets, which should reduce risk whilst maintaining a high expected return.
This idea of efficient diversification is well-founded by Markowitz (1952) in his famous paper \textit{Portfolio Selection}~\cite{Markowitz}.

\subsection{Metrics}
In comparing our algorithm to similar approaches, we use the following metrics that aim to measure performance in a generalised manner. \\
Maximum Drawdown (MDD) \cite{MDD}
\begin{equation}
    \textbf{MDD}(T) = \sup_{\tau \in (0,n)}\left[\sup_{t \in (0,n)} \textbf{S}(t) - \textbf{S}(\tau)\right]
    \label{MDD}
\end{equation}
where $\textbf{S}(\cdot)=\left\{S_{0},S_{1},...,S_{t} \right\}$.

MDD is a risk evaluation metric which represents the maximum decline from a historical peak of the total wealth($S_{i}$) achieved at the time $i$. The smaller the MDD value, the more risk tolerant the trading strategy.

Annualised Percentage Yield (APY) \cite{Elton}
\begin{equation}
    \textbf{APY}_{n} = (S_{n})^{\frac{1}{y}} -1
    \label{APY}
\end{equation}
Here $S_{n}$ is the total return after $n$ trading periods, and $y$ is the number of years corresponding to $n$. APY measures the rate of return that was achieved and it takes into account the effect of compounding.
Typically a greater APY is desired.

Annualised Sharpe Ratio(ASR) \cite{Sharpe}
\begin{equation}
    \textbf{SR}_{n} = \frac{1}{\sigma_{p}}\left( \textbf{APY}_{n} - R_{f} \right)
    \label{ASR}
\end{equation}
Here $\textbf{SR}_{n}$ represents the annualised Sharpe Ratio after $n$ periods, $\textbf{APY}_{n}$ is an Annualised Percentage Yield (Equation~\ref{APY}). $R_{f}$ is the risk-free rate of return and $\sigma_{p}$ is the annualised standard deviation of daily returns.
We use the same assumptions as in the DRICORN-K to calculate Equation~\ref{ASR}~\cite{DRICORN}. Where $R_{f}$ is set to 4\% and $\sigma_{p}$ is set to $\sqrt{252}$ as a result of assuming an average number of 252 trading days in a given year.
The Sharpe Ratio captures the ``return per unit of risk``. A higher value ASR is preferred.

\subsection{Training and Validation}
The data sets used are given in Table~\ref{tab:stockInfo}. Assets had their prices adjusted for dividends and stock splits.  Validation and Testing represent the validation and testing data sets respectively. All sets are in years - where 252 days is the average number of trading days in a year.
The sets consisted of 6000, 5040 and 2520 days for training, validation and testing respectively.

In demonstrating memory's effects, we trained agents using different values of $d$ in intervals of ten between 10 and 230 days. Table~\ref{tab:effects-of-memory} is a subset of results for the best performing sizes for $d$ in various periods. 
\begin{table}[htb!]
    \centering
    \caption{Number of assets in each exchange}\label{tab:stockInfo}
    \begin{tabular}{l|rrrc}
     \toprule
     \bf Exchange & \bf Assets & \bf Validation & \bf Test \\
     \bottomrule\toprule
     BIST & 46 & 2 & 1\\
     BOV & 28 & 4 & 2\\
     EUR & 46 & 4 & 2\\
     JSE & 38 & 4 & 2\\
     NAS & 41 & 4 & 2\\
     SP5 & 47 & 2 & 1\\
     \bottomrule
     
\end{tabular}
\end{table}

\begin{table}[htb!]
    \centering
    \caption{performance of algorithm with various lengths of memory}\label{tab:effects-of-memory}
    \begin{tabular}{l|rrc}
     \toprule
     \bf d & \bf MDD & \bf APY & \bf ASR \\
     \bottomrule\toprule
     10 & -0.277 & 0.665 & 0.028 \\
     120 & -0.300 & 0.806 & 0.034 \\
     190 & -0.252 & 0.723 & 0.030  \\
     \bottomrule
\end{tabular}
\end{table}

It was observed anecdotally that in markets that experienced high volatility with the best performing asset constantly changing, the 10-day agent performed best. In markets that had a consistent best stock over a long period, the 190-day agent performed better. The 120-day agent represents a ``middle of the way``  agent that yielded an overall better strategy as shown by the performance across the presented metrics.

\subsection{Testing}

Each testing data set consists of one year of data, with 300 days prior for CORN-based strategies to train with. This extra data is to allow CORN-based strategies to have a more fair comparison against AMA-K. The CORN-based algorithms were tuned with their optimal hyper-parameters as set out in their respective papers~\cite{CORN, DRICORN, RACORN}.
In the case of our implementation, we have segmented $d$ days for each algorithm, where $d$ is the size of the largest sub-agent's memory - here this would be 190 days. We will compare our approaches to some common baselines such as \textit{UBAH, CRP} and \textit{Best Stock}~\cite{Li}. We also compare our method to EG (Exponential Gradient) as a showcase of a Follow-The-Winner strategy~\cite{EG}. In EG we have set $\eta$ to be $0.05$.

\section{Results and Discussion}
In the Tables~\ref{tab:mdd-1},~\ref{tab:apy} and~\ref{tab:asr}, the number next to a stock exchange represents which data set it came from.
The mean ($\mu$) is the average for the metric across each of the markets and $\sigma$ is the standard deviation.

\begin{table*}[htb!]
    \centering
    \caption{MDD (Maximum Drawdown) of algorithms in various markets.} \label{tab:mdd-1}
    \resizebox{\textwidth}{!}{%
    \begin{tabular}{l|rrrrrrrrrrc}
        \toprule
    	Algorithm        & \bf BIST$^{\downarrow\uparrow}$  & \bf BOV-1$^{\uparrow}$ & \bf EUR-1$^{\downarrow\uparrow}$ & \bf JSE-1$^{\uparrow\downarrow}$ & \bf NAS-1$^{\uparrow\downarrow}$ & \bf SP5$^{\leftrightarrow\uparrow}$ & \bf BOV-2$^{\downarrow\uparrow}$ & \bf EUR-2$^{\uparrow}$ & \bf JSE-2$^{\leftrightarrow\uparrow}$ & \bf NAS-2$^{\downarrow\uparrow}$ & $\mu \pm \sigma$ \\ 
    	\bottomrule\toprule
        
        UBAH        & -0.082    & -0.159   & -0.242    & -0.102    & -0.093    & \bf-0.034 & -0.201    & -0.070 & -0.075 & -0.203 & -0.126 $\pm$ 0.070 \\
        CRP         & \bf-0.081 & -0.150   & -0.248    & \bf-0.097    & \bf-0.090 & -0.035    & -0.199    & -0.068 & -0.072 & -0.204 & -0.124 $\pm$ 0.071\\
        EG          & \bf-0.081 & -0.149   & -0.230    & \bf-0.097    & \bf-0.090 & \bf-0.034 & -0.197    & -0.070 & -0.072 & -0.203 & \bf-0.122 $\pm$ 0.067\\
    	CORN-K      & -0.189    & -0.125   & -0.231    & -0.100    & \bf-0.090 & -0.035    & -0.201    & -0.070 & \bf-0.055 & -0.204 & -0.130 $\pm$ 0.071\\
        RACORN-K    & \bf-0.081 & -0.164   & -0.227    & -0.110    & \bf-0.090 & -0.035    & -0.208    & \bf-0.066 & -0.081 & -0.204 & -0.127 $\pm$ 0.068\\
    	DRICORN-K   & -0.189   & -0.125    & -0.231    & -0.100    & \bf-0.090 & -0.035    & -0.201    & -0.070 & \bf-0.055 & -0.204 & -0.130 $\pm$ 0.071 \\
    	AMA-K       & -0.218   & \bf-0.087    & \bf-0.153    & -0.117    & -0.180    & -0.130    & \bf-0.087    & -0.086 & -0.117 & \bf-0.180 & -0.136 $\pm$ 0.046\\

    	\bottomrule
    \end{tabular}
    }
\end{table*}

\begin{table*}[htb!]
    \centering
    \caption{APY (Annualised Percentage Yield) of algorithms in various markets.}\label{tab:apy}
    \resizebox{\textwidth}{!}{%
    \begin{tabular}{l|rrrrrrrrrrc}
         \toprule
    	Algorithm        & \bf BIST$^{\downarrow\uparrow}$  & \bf BOV-1$^{\uparrow}$ & \bf EUR-1$^{\downarrow\uparrow}$ & \bf JSE-1$^{\uparrow\downarrow}$ & \bf NAS-1$^{\uparrow\downarrow}$ & \bf SP5$^{\leftrightarrow\uparrow}$ & \bf BOV-2$^{\downarrow\uparrow}$ & \bf EUR-2$^{\uparrow}$ & \bf JSE-2$^{\leftrightarrow\uparrow}$ & \bf NAS-2$^{\downarrow\uparrow}$ & $\mu \pm \sigma$ \\
         \bottomrule\toprule
         UBAH & 0.693 & 0.188 & 0.381 & 0.186 & 0.247 & 0.214 & 0.227 & 0.186 & 0.220 & 0.106 & 0.265 $\pm$ 0.166 \\
         CRP & 0.654 & 0.206 & 0.388 & 0.168 & 0.265 & 0.205 & 0.254 & 0.183 & 0.222 & 0.116 & 0.266 $\pm$ 0.154 \\
         EG & 0.640 & 0.203 & 0.421 & 0.167 & 0.280 & 0.204 & 0.253 & 0.185 & 0.220 & 0.131 & 0.270 $\pm$ 0.152\\
         CORN-K & 1.450 & 0.240 & 0.396 & 0.144 & 0.301 & 0.216 & 0.331 & 0.184 & 0.289 & 0.128 & 0.368 $\pm$ 0.390\\
         RACORN-K & 0.484 & 0.124 & 0.302 & 0.107 & 0.288 & 0.219 & 0.169 & 0.143 & 0.215 & 0.117 & 0.217 $\pm$ 0.116 \\
         DRICORN-K & 1.450 & 0.240 & 0.396 & 0.144 & 0.301 & 0.216 & 0.331 & 0.184 & 0.289 & 0.128 & 0.368 $\pm$ 0.390\\
         AMA-K & \bf2.052 & \bf0.539 & \bf1.237 & \bf0.616 & \bf0.418 & \bf0.775 & \bf0.558 & \bf0.390 & \bf0.639 & \bf0.413 & \bf0.764 $\pm$ 0.516\\

         \bottomrule
    \end{tabular}
    }
\end{table*}

\begin{table*}[htb!]
    \centering
    \caption{ASR (Annualised Sharpe Ratio) of algorithms in various markets.}\label{tab:asr}
    \resizebox{\textwidth}{!}{%
    \begin{tabular}{l|rrrrrrrrrrc}
         \toprule
    	Algorithm        & \bf BIST$^{\downarrow\uparrow}$  & \bf BOV-1$^{\uparrow}$ & \bf EUR-1$^{\downarrow\uparrow}$ & \bf JSE-1$^{\uparrow\downarrow}$ & \bf NAS-1$^{\uparrow\downarrow}$ & \bf SP5$^{\leftrightarrow\uparrow}$ & \bf BOV-2$^{\downarrow\uparrow}$ & \bf EUR-2$^{\uparrow}$ & \bf JSE-2$^{\leftrightarrow\uparrow}$ & \bf NAS-2$^{\downarrow\uparrow}$ & $\mu \pm \sigma$ \\
         \bottomrule\toprule
         UBAH & 0.041 & 0.009 & 0.021 & 0.009 & 0.013 & 0.011 & 0.012 & 0.009 & 0.011 & 0.004 & 0.014 $\pm$ 0.010  \\
         CRP & 0.039 & 0.010 & 0.022 & 0.008 & 0.014 & 0.010 & 0.013 & 0.009 & 0.011 & 0.005 & 0.014 $\pm$ 0.010\\
         EG & 0.038 & 0.010 & 0.024 & 0.008 & 0.015 & 0.010 & 0.013 & 0.009 & 0.011 & 0.006 & 0.014 $\pm$ 0.010\\
         CORN-K & 0.089 & 0.013 & 0.022 & 0.007 & 0.016 & 0.011 & 0.018 & 0.009 & 0.016 & 0.006 & 0.021 $\pm$ 0.024\\
         RACORN-K & 0.028 & 0.005 & 0.016 & 0.004 & 0.016 & 0.011 & 0.008 & 0.006 & 0.011 & 0.005 & 0.011 $\pm$ 0.007\\
         DRICORN-K & 0.089 & 0.013 & 0.022 & 0.007 & 0.016 & 0.011 & 0.018 & 0.009 & 0.016 & 0.006 & 0.021 $\pm$ 0.024 \\
         AMA-K & \textbf{0.126} & \bf0.031 & \textbf{0.075} & \textbf{0.036} & \bf0.024 & \textbf{0.046} & \bf0.033 & \textbf{0.022} & \textbf{0.038} & \bf0.023 & \textbf{0.045} $\pm$ \bf 0.032\\
         \bottomrule
    \end{tabular}
    }
\end{table*}
\subsection{Individual Metric Performance}
We examine the performances of the various strategies based on metrics given by Equations~\ref{MDD},~\ref{APY}~and~\ref{ASR}.

\subsubsection{MDD}

As seen in Table~\ref{tab:mdd-1}, the best performing strategies are EG and AMA-K.
EG outperforms AMA-K by a considerable margin on average.
It should be noted that AMA-K achieves a low standard deviation for its MDD.
Despite this lower standard deviation, AMA-K had the greatest MDD value hence, it represents the riskiest strategy.
This is expected given that AMA-K is extremely aggressive in terms of portfolio allocation.
Furthermore, we note that in the two markets that have periods of flatness, we see that AMA-K performed the worst.
AMA-K received an MDD value that is twice and thrice greater than the other strategies (in general) for NAS-1 and JSE-2 respectively.

\subsubsection{APY}

In Table~\ref{tab:apy} we see that AMA-K had the best performance, with AMA-K achieving in the worst case (on NAS-1) 38.87\% better annualised percentage yield for the period compared to the second-best.
AMA-K had the highest mean APY, however it should be noted that its standard deviation is fairly large.
The second-best strategies are CORN-K and DRICORN-K.
These strategies had nearly half the mean APY of AMA-K.
AMA-K's better performance to CORN-K and DRICORN-K results from AMA-K searching for portfolios in days that these CORN-based strategies' experts would have returned uniform portfolios.
AMA-K's strategy has turned enough of these days into profitable trading days - as reflected by its mean APY.

\subsubsection{ASR}

Based on Table~\ref{tab:asr}, we can see that the best performing algorithm is AMA-K.
The performance benefits of AMA-K can be seen in the example of JSE-1, where AMA-K performs 4 times better than the second-best algorithm in JSE-1.
Therefore, we can conclude that per unit of risk with a given risk-free rate of 4\%, the AMA-K algorithm returns showcase the risk to reward trade-off at play.


\begin{figure}[!htb]
    
    \begin{minipage}{\columnwidth}
        \includegraphics[width=1.0\columnwidth]{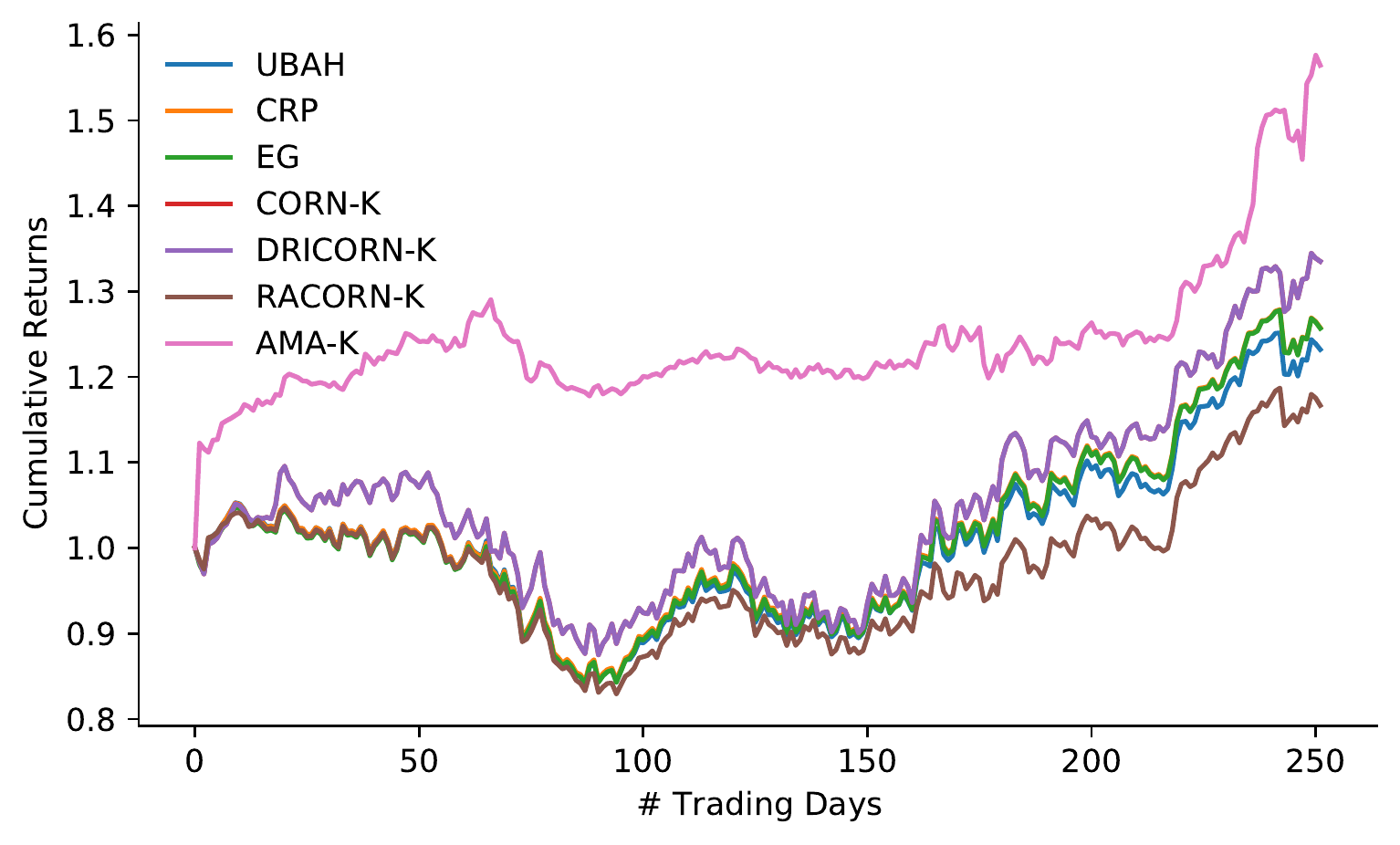}
        \caption{Cumulative returns BOV-2}
        \label{fig:bov-2-down}
    \end{minipage}
    \hfill
    \begin{minipage}{\columnwidth}
    \includegraphics[width=1.0\columnwidth]{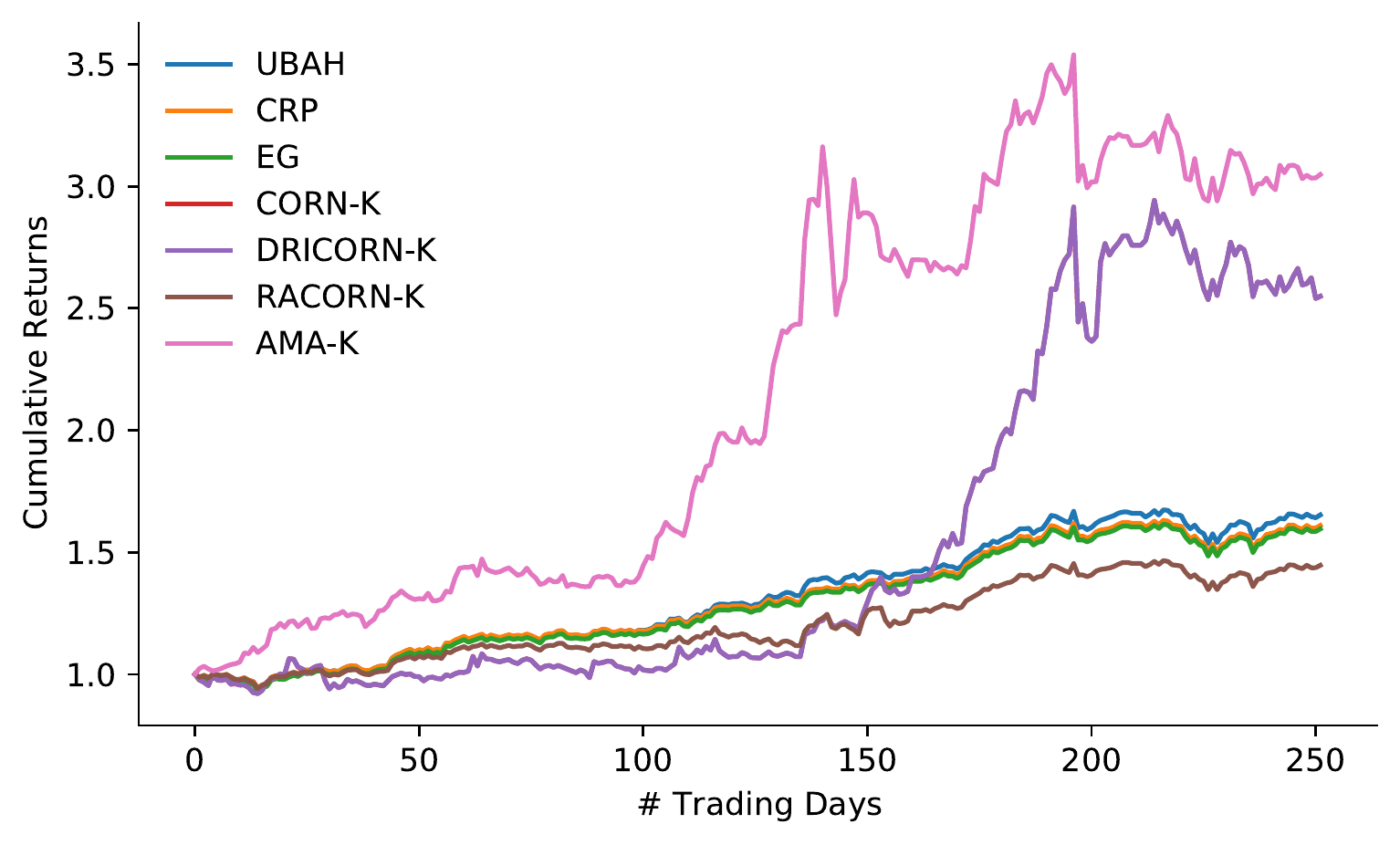}
    \caption{Cumulative returns BIST}
    \label{fig:bist-1-volatile}
    \end{minipage}
    \hfill
    \begin{minipage}{\columnwidth}
    \includegraphics[width=1.0\columnwidth]{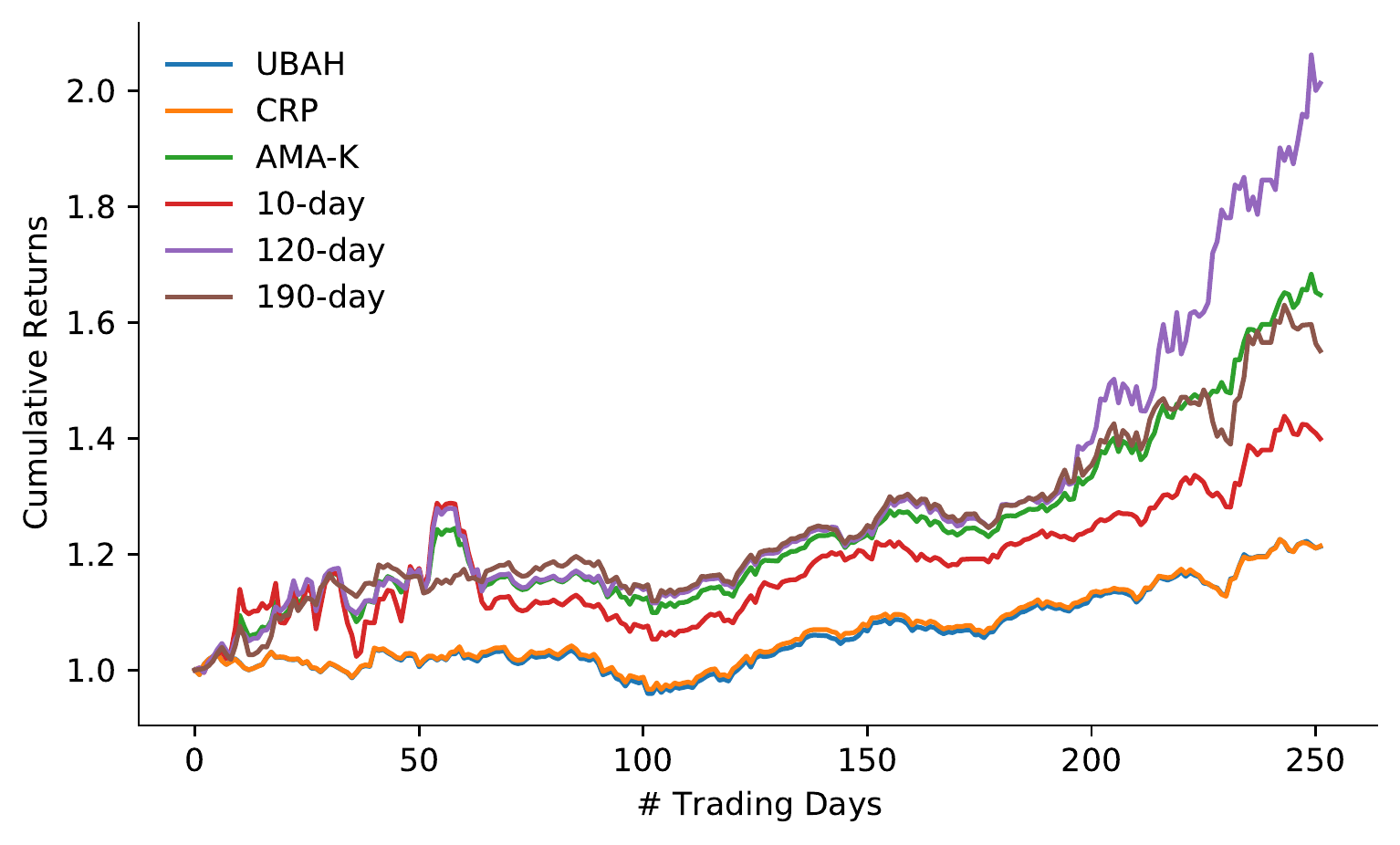}
    \caption{Internal Agents and AMA-K  Cumulative returns JSE-2}
    \label{fig:jse-2-down}
    \end{minipage}
\end{figure}
\begin{figure}[!htb]
    \begin{minipage}{\columnwidth}
        \includegraphics[width=1.0\columnwidth]{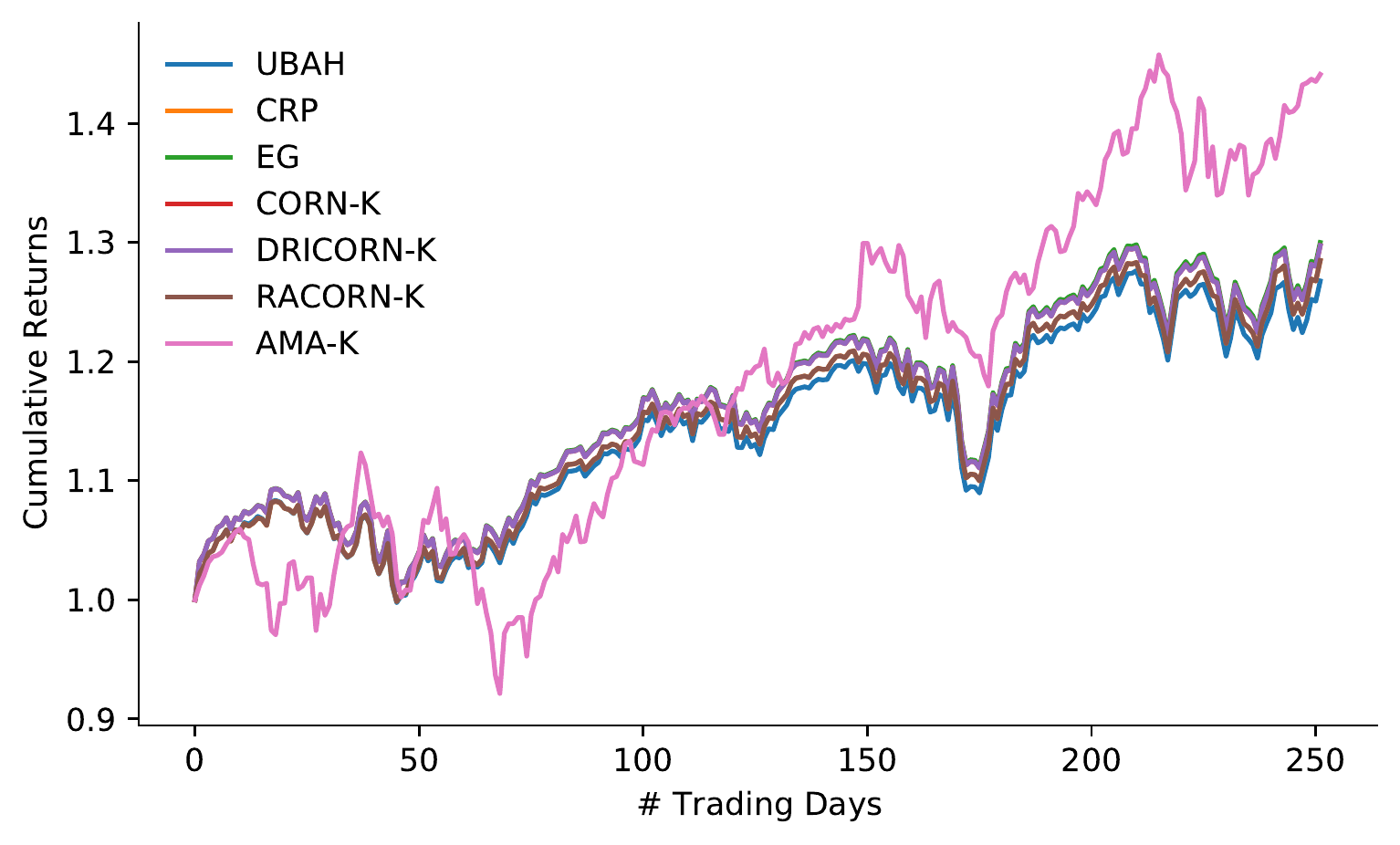}
        \caption{Cumulative returns NAS-1}
        \label{fig:nas-1-up}
    \end{minipage}

\end{figure}

\subsection{General Performance}
Looking at the cumulative return of the algorithm on various markets, we see that the approaches have different patterns in general.
For example for NAS-1 in Figure~\ref{fig:nas-1-up}, the general market trend is upwards, with AMA-K performing the best.
The second-best algorithm is CORN-K and DRICORN-K which have performed the same. For the first 100 days AMA-K's MDD risks are prevalent and AMA-K does the worst. AMA-K fluctuates heavily in this market and from a bottom at day 68 where AMA-K lost 7.8\% of its total wealth in four days to being the best performing algorithm at the end.

In Figure~\ref{fig:jse-2-down} we see the market stays relatively flat before decreasing slightly for the first 100 days. Comparing our 10-day, 120-day and 190-day sub-agents to AMA-K we see that AMA-K cannot beat the best performing agent. AMA-K's other agents have also done well for the period and for the initial 55 days we see that the sub-agents perform well. The 10-day agent is clearly more volatile and loses more of its cumulative wealth than the other agents. Towards the end of the period, the 120-day agent has out-performed all other strategies. AMA-K has benefited from this sub-agent, but the benefit is dampened by the other sub-agents. 

When CORN-K's and DRICORN-K's experts pick up enough similarity, their performance is close to ours.
This can be seen in Figure~\ref{fig:bist-1-volatile}.
Despite our approach making considerable gains early on, the CORN-K and DRICORN-K approaches identified an asset our approach was unable to.
This resulted in these algorithms far outpacing our method.
The asset that CORN-K and DRICORN-K identified was most likely in a period our agents had forgotten.

In general our approach is competitive and leverages CORN-based strategies to produce further gains by searching for optimal portfolios on days that CORN-based strategies would not. Although the risk presented by our algorithm as shown in Figure~\ref{fig:nas-1-up} can be significant, it is an example of a risk to reward trade-off.

\section{Conclusion and Further Development}
Our approach can generate high returns using our memory-based method.
The approaches can be volatile and merging them typically results in a more stable strategy (as shown in Tables~\ref{MDD},~\ref{APY},~\ref{ASR}) at the cost of reducing returns of the best agents.
Here are possible areas to develop further:
Changing how we search for the optimal portfolio to include regret or penalise volatility - for lower-risk strategies.
Doing a fine-grained grid search for the optimal amount of memory for the agents in different periods.
Testing AMA-K using a variety of combinations for the $d$-day agents with different time horizons.
Lastly, further testing using other well known or niche financial metrics should be conducted to further understand the performance of the method in comparison to other modern approaches.

\bibliographystyle{IEEEtran}
\bibliography{amak}

\end{document}